%% file: article.tex
\documentclass[aps,prc,showpacs,twocolumn,floatfix,nobibnotes,nofootinbib,amsmath,amssymb,longbibliography]{revtex4-2}
\include{preamble}

\begin{document}

\title{Posterior predictive distributions of neutron-deuteron cross sections}

\date{\today}
\author{Sean B.\ S.\ Miller}
\author{Andreas Ekstr\"om}
\author{Christian Forss\'en}
\affiliation{Department of Physics, Chalmers University of Technology, SE-412 96 Gothenburg, Sweden}

\begin{abstract}
We quantify the posterior predictive distributions (\ppd{}s) of
elastic neutron-deuteron (\nd{}) scattering cross sections using
nucleon-nucleon (\NN{})
interactions from chiral effective field theory (\cheft{}) up to and
including next-to-next-to-next-to-leading order (\NNNLO). These
\ppd{}s quantify the spread in \nd{} predictions due to the
variability of the low-energy constants (\lec{}s) inferred from \NN{} scattering data. We use
the wave-packet continuum discretization method to solve the
Alt-Grassberger-Sandhas form of the Faddeev equations for elastic
scattering. We draw 100 samples from the \ppd{}s of \nd{} cross
sections up to 67 MeV in scattering energy, i.e., in the energy region
where the effects of three-nucleon forces are expected to be small. We
find that the uncertainty about \NN{} \lec{}s inferred from \NN{}
scattering data, when assuming uncorrelated errors, does not translate
to significant uncertainty in the low-energy \nd{} continuum. Based on
our estimates, the uncertainty of \nd{} predictions are dominated by
the \cheft{} truncation error, at least below \NNNLO{}. At this order,
the 90\% credible interval of the \ppd{} and the truncation error are
comparable, although both are very small on an absolute scale.
\end{abstract}

\maketitle

\section{Introduction}
\label{sec | introduction}
Chiral effective field theory
(\cheft{})~\cite{Epelbaum:2008ga,Machleidt:2011zz,Hammer:2019poc}
promises a systematically improvable description of the nuclear
interaction grounded in the symmetries of low-energy quantum
chromodynamics. Two-nucleon (\NN{}) and three-nucleon (\NNN{})
interactions from \cheft{} are used extensively in modern \textit{ab
  initio} predictions of atomic nuclei and nuclear matter, see, e.g.,
Refs.~\cite{Hergert:2020bxy,Hebeler:2020ocj,Tews:2020hgp} for recent
overviews. To make quantitative predictions of the properties of
nuclear systems, the numerical values of the low-energy constants
(\lec{}s) that govern the strengths of the pion-nucleon (\piN) and
nucleon-contact couplings must first be inferred from
low-energy data. For this, the Bayesian approach to
statistics~\cite{bda3} provides a natural framework since it yields a
(posterior) probability density function (\pdf{}) that quantifies our
uncertainty about the values of the \lec{}s. Propagating this
uncertainty when making theoretical predictions amounts to
averaging the distribution of predictive samples over the \lec{}
posterior \pdf{}. The result of this is called a posterior predictive
distribution (\ppd{}). This type of distribution sits at the center of
the scientific process whereby we try to predict future data based on
previous data and theory.

There are existing efforts to quantify Bayesian \ppd{}s for various
nuclear observables, e.g., \NN{} scattering cross
sections~\cite{Svensson:2021lzs} and scattering
lengths~\cite{Svensson:2022kkj}, few-nucleon~\cite{Wesolowski:2021cni}
and many-nucleon~\cite{Djarv:2021pjc,Stroberg:2019bch,Hu:2021trw}
energies, radii, and decays, as well as nuclear mass
models~\cite{Neufcourt:2019qvd,Kejzlar:2020vla}. These probability
distributions quantify our degree-of-belief, and facilitate a
meaningful comparison with experimental data.
For example, the \ppd{} finds
use in model checking~\cite{bda3}, such as posterior predictive checks.
There one simulates data, using a fitted model, and compares to
observed data. The simulated data corresponds to draws from the \ppd{}
and it should look roughly like the observed data if the model did indeed
contain all relevant physics and there has been a sufficient amount of
calibration data.

In this work, we sample the \ppd{}s of selected neutron-deuteron
(\nd{}) scattering cross sections arising from the variability of the
\lec{} posterior when conditioned on \NN{} scattering data. We use \cheft{}
descriptions of the \NN{} interaction at all orders up to
next-to-next-to-next-to-leading order (\NNNLO{}) in Weinberg power
counting. To the best of our knowledge there exists only
frequentist statistical analyses encompassing a subset of nucleon-deuteron (\Nd{})
scattering cross sections and scattering
lengths~\cite{LENPIC:2018ewt,Skibinski:2018dot,Volkotrub:2020lsr}, for
which various estimates for dispersion have been quantified.
Our analysis is rooted in Bayesian methodology and therefore
provides probability densities for the predicted observables of
interest. As such, the results
of this work facilitates a quantitative measure of the predictive
power in the low-energy \NNN{} continuum using \cheft{} interactions
carefully calibrated using \NN{} scattering data. This work is part of
an ongoing effort towards a full Bayesian analysis of \cheft{}
conditioned also on experimental data in the \Nd{} continuum
\cite{Miller:2022beg}.

To sample the \ppd{}s of elastic \nd{}-scattering cross sections, we
repeatedly solve the Alt-Grassberger-Sandhas~\cite{Alt:1967fx} (AGS)
form of the Faddeev equations using the wave-packet continuum
discretization (\wpcd{})
method~\cite{Rubtsova:2015owa,Miller:2022beg}. This method is
parallelizable with respect to the scattering energy, denoted
with $\Elab$, in the laboratory frame of reference. Therefore, it is
particularly suitable for sampling \ppd{}s across a range of $\Elab$
values. Still, the collection of samples from the \ppd{}s is
limited by the number of times we can solve the AGS equation. For this reason, we
currently neglect \NNN{} forces (\NNNF{}s) and focus
our analysis on cross sections and polarization observables with
$\Elab \leq 67 $ MeV, for which \NN{}-only models typically perform
well~\cite{Witala:2000am,Kalantar-Nayestanaki:2011rzs,Epelbaum:2019zqc,LENPIC:2018ewt}.
The low-energy vector analyzing power, $A_y$, is a possible exception
to this statement and we therefore place a special
focus on the analysis of this polarization observable.

In addition to the inherent uncertainty of inferred \lec{} values,
there are also other sources of theoretical uncertainty. The model discrepancy due to the
omission of higher chiral orders is an obvious one. Neglecting this
uncertainty can lead to biased and over-confident inferences and
predictions~\cite{Brynjarsd_ttir_2014}. Fortunately, \cheft{} is
designed to be an order-by-order improvable description of the nuclear
interaction, and as such the theory itself provides valuable
information about the magnitude of the truncation
error~\cite{Furnstahl:2015rha}. Indeed, there exists several efforts
to quantify the truncation error in effective field theory predictions
of nuclear systems, see, e.g.,
Refs.~\cite{CoelloPerez:2015ksi,Ekstrom:2017koy,Hammer:2017tjm,Melendez:2017phj,Drischler:2020hwi,Wesolowski:2021cni,Svensson:2021lzs}.
Although our focus is to quantify the \ppd{}s of \nd{} scattering
observables due to variability in the
\NN{} \lec{}s, we will also contrast our findings with estimates of the truncation
error.

In Sec.~\ref{sec | bayesian methods} we define the general structure
of the \ppd{}s we sample in this work. In Sec.~\ref{sec | wpcd} we
present the essential elements of the \wpcd{} method we use to produce
elastic \nd{} cross sections. In Sec.~\ref{sec | results} we present
the strategy for sampling the \ppd{}s, with particular focus on
achieving computational speedup, and the results of the sampling. We
also compare the degree-of-belief intervals of the \ppd{}s with some
of the other components of the total error budget; the
\cheft{} truncation error in particular. We end with a summary and outlook in
Sec.~\ref{sec | summary and outlook}.

\section{Setting up the posterior predictive distribution}
\label{sec | bayesian methods}
The \ppd{} is a \pdf{} $\pr(y|D,M,I)$ for a quantity $y$ as predicted
by a model $M$. This distribution quantifies the uncertainty about $y$ 
given previous data $D$ and any other assumptions or information $I$. Here, 
we focus on the uncertainty of the numerical values of the \lec{}s, denoted
$\lecs$, present in the underlying \cheft{} \NN{} interaction. As such,
we must marginalize over the \lec{}s by evaluating the following integral
\begin{align}
  \begin{split}
  \label{eq | ppd}
  \pr(y|D,M,I) {}& = \int_{\Omega} \pr(y|\lecs,D,M,I)\pr(\lecs|D,M,I)\,d\lecs \\ {}& \propto \int_{\Omega} y(\lecs) \pr(\lecs|D,M,I)\,d\lecs.
  \end{split}
\end{align}
In the second line we introduced a short-hand $y(\lecs)$ for a
deterministic model prediction given numerical values for $\lecs$ from 
some parameter domain $\Omega$. We also used that $y$ is conditionally 
independent of $D$. The proportionality indicates that we are only 
interested in the width and shape of the \ppd{}, and not the overall 
normalization constant.

We will refer to the \cheft{} description of the \NN{} interaction at
a chiral order $\nu$ as a ``model'', and denote this as $M_{\nu}$. The
chiral orders are defined according to Weinberg power counting with
$\nu=0,2,3,4$, and as is common, we refer to them to as leading order
(\LO{}), next-to-leading order (\NLO{}), next-to-next-to-leading order
(\NNLO{}), and \NNNLO{},
respectively. The values of $\lecs$ depend on the chiral order
$\nu$, but to simplify notation we do not index $\lecs$ by $\nu$.

The \ppd{} is a probabilistic generalization of the familiar
point-estimate value $y_{\star} = y(\lecs_{\star})$, obtained by
evaluating the model $M_{\nu}$ at some preferred parameter value
$\lecs_{\star}$, such as a local parameter-optimum. We will in some
cases resort to evaluating the \ppd{} at the \textit{maximum a
  posteriori} (\map{}) value of the \lec{} posterior
\begin{equation}
  \lecs_{\star} \equiv \underset{\lecs}{\mathrm{argmax}} \,\,\pr(\lecs|D,M_\nu,I).
\end{equation}
Note that the \ppd{} does not necessarily attain its maximum for
$\lecs_{\star}$. Indeed, the evaluation of $y(\lecs)$, through the AGS
equation, is neither linear nor monotonic.

Evaluating the integral in Eq.~\eqref{eq | ppd} requires knowledge about the \pdf{}, $\pr(\lecs|D,M_{\nu},I)$. We utilize the available \lec{}
posteriors up to and including \NNNLO{} published in
Ref.~\cite{Svensson:2022kkj}. These posteriors were sampled using
Hamiltonian Monte Carlo (\hmc) while accounting for uncorrelated
\cheft{} truncation errors, and were conditioned on the Granada
database~\cite{NavarroPerez:2013mvd,NavarroPerez:2013usk} of \NN{}
scattering cross sections for scattering energies $\Elab\leq 290$
MeV. The leading neutron-neutron (\nn) isospin-breaking \lec{} was
inferred using an empirical value for the \nn{} scattering length in
the $^{1}S_0$ partial-wave channel. We note that other methods accounting for correlated \cheft{} truncation errors exist, see, e.g.,  Ref.~\cite{Melendez:2019izc}, which may change the inferred, and rather narrow, distributions of \lec{} values we use here.

The \hmc{} algorithm is particularly well-suited for sampling high-dimensional \pdf{}s
and yields virtually uncorrelated draws from $\pr(\lecs|D,M_{\nu})$. A
detailed analysis~\cite{Svensson:2021lzs} suggests that the \hmc{} chains
we employ in this work to represent the \lec{} posteriors are
sufficiently converged at all orders, unimodal, and rather
concentrated in parameter space. As such, we have in-depth knowledge of
the location of the posterior mass, which helps tremendously when
evaluating the integral in Eq.~\eqref{eq | ppd}. 

\section{Wave-packet continuum discretization}
\label{sec | wpcd}
In this section we summarize the \wpcd{}
method~\cite{Rubtsova:2015owa} for solving the AGS equation in
momentum space. Our results are based on the implementation presented
in Ref.~\cite{Miller:2022beg}\footnote{The implementation, named
``Tic-tac'', is available under a GNU open-source license (GPLv3) on
\href{https://github.com/seanbsm/Tic-tac}{\texttt{https://github.com/seanbsm/Tic-tac}}}.
The AGS equation for \nd{} scattering, without \NNNF{}s, can be written as
\begin{equation}
	\hat{U} = \hat{P}\hat{G}_0^{-1} + \hat{P}\hat{t}_1\hat{G}_0\hat{U} \:,
	\label{eq | AGS standard}
\end{equation}
where $\hat{U}$ is the transition matrix between asymptotic scattering
states, $\hat{G_0}\equiv\frac{1}{E-\hat{h}_0\pm i\epsilon}$ is the
resolvent of the free \NNN{} Hamiltonian $\hat{h}_0$, $E$ is the total energy, $\hat{t}_1$ denotes the scattering $T$-matrix for
the pair-system $(23)$ as written in standard odd-man-out notation,
and $\hat{P}\equiv2\hat{P}_{123}$ where $\hat{P}_{123}$ is the
permutation matrix acting on partially-antisymmetric \NNN{}
states\footnote{There is an erroneous extra term $+1$ in the
definition of $\hat{P}$ in Ref.~\cite{Miller:2022beg}.}. The large
dimensionality of the \NNN{} Hilbert space makes it challenging to
apply matrix-inversion type methods to solve Eq.~\eqref{eq | AGS
  standard}. Instead, one usually resorts to expanding the AGS
equation in a Neumann series that is subsequently re-summed
using a Pad\'e approximant \cite{george1975essentials} to handle the
divergence originating from the integral-kernel $\hat{G}_0\hat{v}_1$
with Weinberg eigenvalues~\cite{Weinberg:1963zza} outside the unit
circle.

It is well understood how to obtain converged solutions for $U$ in a
standard plane-wave basis, see, e.g., Ref.~\cite{Gloeckle:1995jg}. In
this basis, $\hat{U}$ is obtained for a specific value of the on-shell
energy $E$, and the resolvent $\hat{G_0}$ and \NN{} $T$-matrix
$\hat{t}_1$ depend explicitly on $E$. This dependency inflicts several
complications such as moving singularities in the resolvent operator,
and a requirement for antisymmetrizing \NN{} $T$-matrices at many
energies when evaluating the AGS integral kernel, which is typically
handled using splines~\cite{Glockle:1982agg}.

In this work, we use the \wpcd{} method~\cite{Rubtsova:2015owa} for
solving the AGS equation. This is one of many bound-state
approaches~\cite{Carbonell:2013ywa} for describing scattering
processes. In~\wpcd{}, we discretize the continuum using a wave-packet
basis. Doing so simplifies the numerical analysis of the AGS
equation. First, one can derive a closed-form expression of the
channel-resolvent, treating the associated singularities
analytically. Second, the $P$-matrix has no need for splining.
Third, it factorizes the on-shell energy dependence out
of the matrix multiplications associated with the terms of the Neumann
series expansion, providing significant speedup of the most
time-consuming parts of the numerical solution.

As a downside, the \wpcd{}
method entails large matrix dimensionalities compared with the
plane-wave representation. However, scattering amplitudes can be calculated at multiple
scattering energies with minor extra computational cost per energy. This makes
\wpcd{} particularly suitable for sampling Bayesian \ppd{}s across
ranges of energies. In fact, we find that calculating
scattering amplitudes at multiple scattering energies only doubles the
computational cost compared to computing the amplitude at a single
energy~\cite{Miller:2022beg}

We define a wave packet $|x\rangle$ as a finite integral of continuum
states $|p\rangle$, e.g., plane-wave states, within a momentum ``bin''
$\mathcal{D}\equiv[p,p+\Delta p]$,
\begin{equation}
	|x\rangle \equiv \frac{1}{\sqrt{N}}\int_\mathcal{D} f(p') |p'\rangle \:p'\:dp' \:,
\end{equation}
where $f(p)$ is a weighting function and $N$ is the normalization
constant. An $A$-body wave packet can be straightforwardly defined using wave-packet discretization for
each Jacobi coordinate. A \NNN{} wave packet is given by
$|X\rangle\equiv|x\rangle\otimes|\bar{x}\rangle$, where $|x\rangle$
corresponds to the pair-system $p$-momentum and $|\bar{x}\rangle$
corresponds to the spectator $q$-momentum.

The eigenstates of the \NN{} Hamiltonian $\hat{h}_1$ in a (plane-wave)
wave-packet basis can be used to approximate ``scattering'' \NNN{}
wave packets rather well. In this basis, it is also possible to
evaluate the channel-resolvent $\hat{G}_1\equiv
\frac{1}{E-\hat{h}_1\pm i\epsilon}$ analytically. Furthermore, using
that $\hat{t}_1\hat{G}_0 \equiv \hat{v}_1\hat{G}_1$ and
$\hat{G}_0^{-1} = \hat{v}_1$ (on-shell), we can rewrite Eq.~\eqref{eq |
  AGS standard} to obtain
\begin{equation}
	\hat{U} = \hat{P}\hat{v}_1 + \hat{P}\hat{v}_1\hat{G}_1\hat{U}
        \:,
	\label{eq | AGS}
\end{equation}
where $\hat{U}$ now depends on $E$ only via $\hat{G}_1$. This is the
starting point for solving the AGS equation in the \wpcd{}
method. Here, as in Ref.~\cite{Miller:2022beg}, we use an equal number
of wave packets, $\nwp$, to discretize the $p$ and $q$ continua,
yielding matrices in Eq.~\eqref{eq | AGS} that scale in size as
$\mathcal{O}(\nwp^4)$. We find that the runtime of the code follows
this quartic scaling with $\nwp$ quite closely. Note, however, that
the calculations at \NNNLO{} are $\sim 10\%$ more costly since the
Pad\'e resummation of the Neumann series typically requires more terms
to converge.

\section{Evaluating posterior predictive distributions}
\label{sec | results}
We sample the \ppd{} of a scattering observable by evaluating Eq.~\eqref{eq | ppd} numerically. This is done by computing the \nd{}
scattering observable of interest for a finite set of \lec{} values
drawn from the posterior \pdf{}, $\pr(\lecs|D,M_{\nu})$. In practice, we use the
Markov chains obtained in Ref.~\cite{Svensson:2022kkj}.

For every
sample that we draw from the \ppd{} we must solve the AGS
equation. Fortunately, with the \wpcd{} method we get access to all
scattering cross sections at all angles and energies without any
significant computational overhead. Also, since the permutation
operator $\hat{P}$ does not depend on the \lec{}s, we only have to compute
this once and re-use it throughout the sampling process. However, we
have to setup the Neumann series for every new sample, and this is the
most time-consuming part.

In all calculations done here, we use a spin-angular basis of \NNN{}
partially-antisymmetric partial-waves with total angular momentum
$\mathcal{J} \leq 17/2$, using both parities, and using \NN{} total
angular momentum $J\leq 3$. We also explicitly account for the charge
dependence of the strong \NN{} interaction in the $^1$S$_0$
channel. This state space provides sufficiently converged $U$-matrix
elements for $\Elab \leq 100$ MeV when using the chiral potentials
defined in Ref.~\cite{Svensson:2021lzs,Svensson:2022kkj}.  Note that
our study is limited to $\Elab\leq 67$ MeV due to the omission of
\NNNF{}s. It has been shown that, at low scattering energies, the
scattering amplitudes are likely dominated by \NN{}
forces~\cite{Witala:2000am,Kalantar-Nayestanaki:2011rzs,Epelbaum:2019zqc,LENPIC:2018ewt}.

We discuss our general strategy to quantify
the \ppd{} in Sec.~\ref{sec | convergence}, present results for the
\ppd{}s of the differential \nd{} cross section in Sec.~\ref{sec |
  dsg}, relate this to estimates of the \cheft{} truncation errors in
Sec.~\ref{sec | eft error}, and discuss spin-polarization observables,
focusing on $A_y(n)$, in Sec.~\ref{sec | polarization}.

\subsection{Trading wave-packets for computational speedup}
\label{sec | convergence}
In the limit $\nwp\rightarrow\infty$, the \wpcd{} results converge
towards the results from an exact calculation, e.g., a continuous
plane-wave solution~\cite{Gloeckle:1995jg} of Eq.~\eqref{eq | AGS}. 
However, the computational cost increases
quartically with $\nwp$, and larger values for $\nwp$ will significantly
increase the \ppd{} sampling cost. Balancing cost and accuracy, we found it sufficient to
draw $N=100$ samples from each \ppd{} that we study, since we are
quantifying univariate distributions. Also, we noticed
that the shapes and widths of the \ppd{}s studied here did not change
visibly when varying $\nwp$, and as such we could limit ourselves to
$\nwp \leq 75$ and extrapolate to larger values. This will be
discussed in the next section.

At present, using $\nwp=75$, it takes roughly 12 node-hours (384
core-hours\footnote{Using two Intel Xeon Gold 6130 CPUs per node,
amounting to 32 cores per node.}) to compute all necessary scattering
amplitudes at $\sim$50 scattering energies below $100$ MeV for a
single configuration of values for the \lec{}s at a specific chiral
order $\nu$. This translates to roughly 150k core-hours to compute all
scattering amplitudes for 100 different \lec{} values at four chiral
orders. The same calculation with $\nwp=150$ would be 16 times more
expensive and cost roughly 2.5M core-hours. To monitor the reduced method
accuracy at $\nwp=75$, we repeat the \ppd{}
sampling, with copies of the same \lec{} samples, at every chiral
order with $\nwp=30$ and $50$. We also use a restricted set of 10
posterior samples with
$\nwp=100$. In addition, we evaluate the \ppd{} at the \map{} value
$\lecs_{\star}$ of the \lec{} \pdf{} using $\nwp=30,\:50,\:75,\:100$,
and $150$. The $\nwp=75,150$ \map{} predictions will be used for extrapolation in the next section.

We had little cost-related reason to restrict calculations to $\Elab\leq 67$ MeV. Instead, we computed the on-shell
$U$-matrices at all wave-packet \NN{} Hamiltonian eigenenergies below
$\Elab=100$ MeV, which was roughly two thirds of the wave-packet basis
size.  Between these energies we perform linear interpolation of the
$U$-matrix elements to virtually any $\Elab<100$ MeV.  Consequently,
we obtained 100 samples from the \ppd{} of any
elastic scattering cross section at every order up to, and including,
\NNNLO{}. Of course, with the neglect of \NNNF{}s, we consider our predictions
above $\Elab=67$ MeV to be incomplete and have therefore been omitted from the present study.
Nonetheless, they allowed us to check on the width and shape of \ppd{}s
all the way to $\Elab=100$ MeV.

\begin{figure}[t!]
  \centering
  \includegraphics[width=\columnwidth]{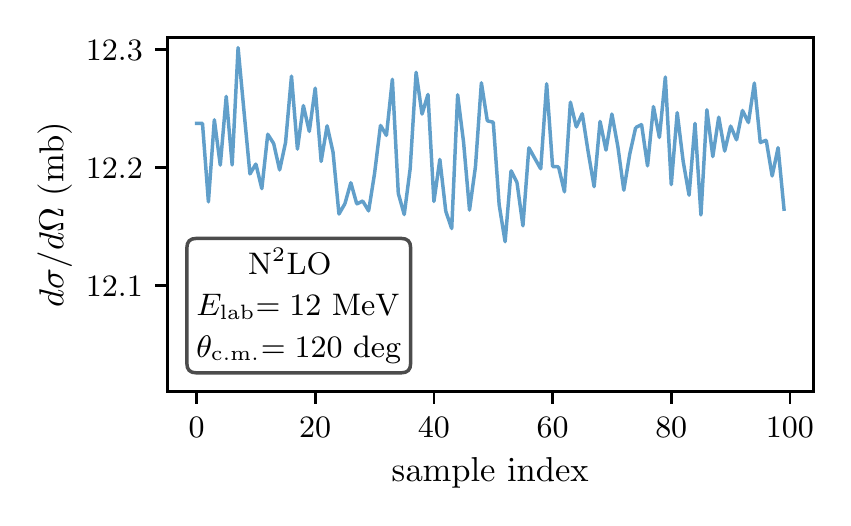}
  \caption{Trace plot of the differential cross section \ppd{} at
    \NNLO{}, for $\Elab=12$ MeV and $\thcm=120$ degrees, using 100
    samples from the \hmc{}-chain of samples from the \lec{} posterior
    at this order.}
  \label{fig | ppd trace}
\end{figure}

Although the \hmc{}-chains of \lec{} posterior samples are virtually
uncorrelated, this does not imply that ensuing samples from the \nd{}
cross section \ppd{} are equally uncorrelated. Unfortunately, a chain
of 100 samples is typically too short to quantify, e.g., an integrated
autocorrelation time or reliably determine the autocorrelation
itself. Nevertheless, an inspection of the trace plots of the \ppd{}
samples, as shown in Fig.~\ref{fig | ppd trace}, does not indicate any hints of strong correlation between
samples.
In the event of observing strongly correlated samples, the information
content of the \ppd{} chain, as measured by its effective sample size,
will drop inversely to the integrated autocorrelation time and we
would have to increase the number of samples
accordingly~\cite{Svensson:2021lzs}

\subsection{The differential cross section}
\label{sec | dsg}
The convergence of the differential elastic \nd{} cross section at
$\Elab = 12$ MeV with respect to $\nwp$ is shown in Fig.~\ref{fig |
  nwp convergence}.
\begin{figure}[t!]
  \centering
  \includegraphics[width=\columnwidth]{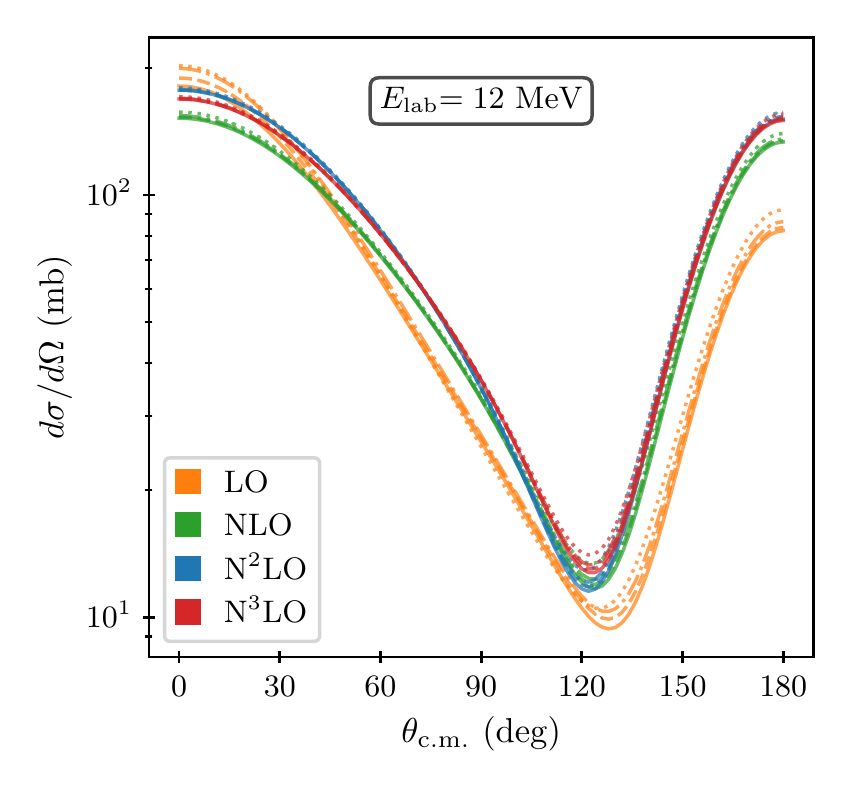}
  \caption{The differential \nd{} cross section at $\Elab=12$ MeV
    computed using the \map{} values for the \lec{}s at chiral orders from \LO{} to
    \NNNLO{}. The dotted, dash-dotted, dashed, and solid lines at each
    order show the results obtained from the \wpcd{} methods with
    $\nwp=50,\:75,\:100$, and $150$, respectively.}
  \label{fig | nwp convergence}
\end{figure}
Clearly, with $\nwp \approx 100$, the results begin to stabilize, at
least for subleading orders. The somewhat reduced convergence rate
for the \LO{} results might be caused by the rather coarse wave-packet
representation of the \NN{} potential for low relative momenta
\cite{Miller:2021vby}. To remedy this one should either re-distribute
the discretization boundaries to improve the coverage of the lower
momentum region, or simply increase $\nwp$ if possible. Since we
detect a sufficient convergence at subleading orders, we opt for
keeping the discretization mesh the same throughout all calculations
and at all chiral orders.

\begin{figure}[t!]
  \centering
  \includegraphics[width=\columnwidth]{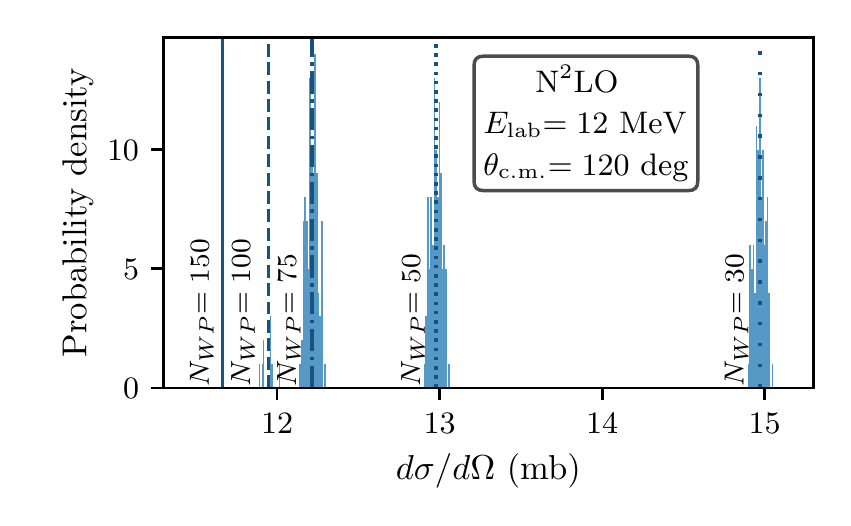}
  \caption{The \ppd{}s of the differential \nd{} cross section at
    $\thcm=120$ degrees and $\Elab=12$ MeV using the \NNLO{} \NN{}
    interaction. The three different distributions shown are, from
    left to right, for $\nwp=100,\:75,\:50$, and 30, with
    $N=10,\:100,\:100$, and 100 samples, respectively. For comparison,
    we also indicate with vertical lines the locations of the cross
    section for the \map{} \lec{} point obtained in a \wpcd{}
    calculation based on $\nwp=150,\:100,\:75,\:50$, and 30. The means
    of the distributions coincide almost with the \map{} predictions.}
  \label{fig | ppd nwp}
\end{figure}

Next, we study the convergence of the \ppd{} with respect to
$\nwp$. In Fig.~\ref{fig | ppd nwp} we show a histogram of 100 samples
of the \ppd{} of the \nd{} differential cross section at $\Elab=12$
MeV and $\thcm=120$ degrees at \NNLO{} using $\nwp=30,\:50$, and 75,
as well as 10 samples at $\nwp=100$ and the location of the \map{}
prediction using $\nwp=150$.
The \ppd{}s based on $\nwp=30,\:50$, and $75$ are very similar in
terms of shape and width. In fact, for all observables that we study
in this work\footnote{We study the differential cross section,
$\frac{d\sigma}{d\Omega}$, the neutron vector analyzing power,
$A_y(n)$, and the spherical tensor analyzing powers $iT_{11}$,
$T_{20}$, $T_{21}$, and $T_{22}$, at angles $\thcm=60$ and 120 degrees
and scattering energies $\Elab=10$-$12,\:35$-$36$, and 65-67 MeV.}, the width and
shape of the \ppd{} remains approximately constant as we vary $\nwp$,
and the main effect is a shift of the entire distribution. Therefore, we
shift the mean of the samples obtained with $\nwp=75$
using the difference between the \map{} predictions obtained with
$\nwp=75$ and $\nwp=150$. This makes a comparison with experimental
data more meaningful.

We did not detect a robust exponential or
power-law convergence pattern with respect to $\nwp$ and leave further
analysis of the $\nwp$-convergence and the \wpcd{} method uncertainty
to future work. As such, there might be additional corrections to the
\ppd{}s when using $\nwp>150$ that we do not account for. However,
assuming that the widths and shapes of the \ppd{}s remain unchanged,
our main conclusions in this work will not be affected.

\begin{figure}[t!]
  \centering
  \includegraphics[width=\columnwidth]{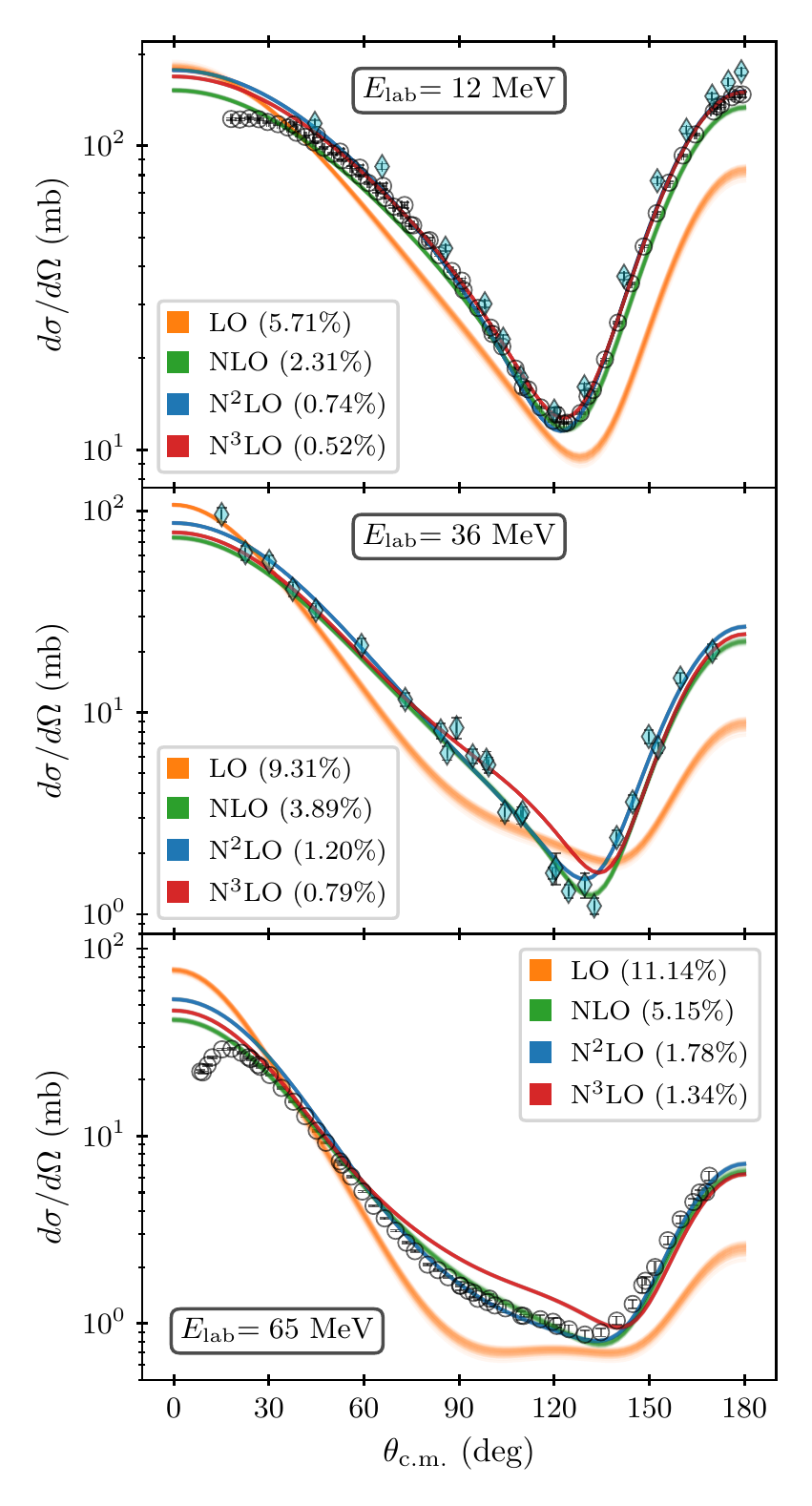}
  \caption{The \ppd{} of the \nd{} differential cross section at
    $\Elab=12,\:36$, and 65 MeV at all orders up to \NNNLO{} in the
    \NN{} interaction. The legends display the average 90\% credible
    intervals (\hpdi{}), see
    text for details. The experimental data (markers) are retrieved
    from the EXFOR database (at $\Elab=12$ and 36 MeV) and
    Ref.~\cite{Shimizu:1982lqa} (at $\Elab=64.5$ MeV). Diamond (cyan) and
    circle (colorless) markers represent \nd{} and \pd{} cross sections,
    respectively.} 
  \label{fig | ppd dsg}
\end{figure}

After shifting the differential cross section obtained with $\nwp=75$
to $\nwp=150$ we obtain the result shown in Fig.~\ref{fig | ppd dsg}.
At all chiral orders and energies we study, the \ppd{} is rather
narrow. At \LO{}, the \ppd{} width is comparable to the experimental
uncertainty, while at subleading orders the experimental uncertainty is
typically greater than the width of the \ppd{}.

To quantify the width
of the \ppd{}s, we compute the 90\% highest posterior density interval
(\hpdi{}), normalize it to the mean of the \ppd{}, and average over
$\thcm$. This way, we find that the average \hpdi{} for the
differential cross section at $\Elab=12$ MeV is, 5.7\%, 2.3\%, 0.7\%,
and 0.5\% at \LO{}, \NLO{}, \NNLO{}, and \NNNLO{}, respectively. The
decreasing values reflects the increasingly narrow \lec{} posterior
densities obtained at higher chiral
orders~\cite{Svensson:2021lzs,Svensson:2022kkj}.  Moving to higher
scattering energies we find that the \ppd{}s remain very narrow still.
Apart from \LO{}, the average \hpdi{} values are comparable to
frequentist estimates of dispersion quantified in
Ref.~\cite{Skibinski:2018dot}, where a similar increase in uncertainty
was noted at higher scattering energies.

Recently it was shown that \NNLO{} potentials with \NNNF{}s yield an
excellent description of differential cross section
data~\cite{Epelbaum:2019zqc}. It was suggested in
Ref.~\cite{Witala:1998ey} that \NNNF{}s are necessary to reproduce the
differential cross section minimum in the vicinity of $\Elab=65$
MeV. Here, however, we see similar reproduction of data at \NLO{} and
\NNLO{} without \NNNF{}s. Going to \NNNLO{}, the reproduction of
experimental data deteriorates. As shown in
Ref.~\cite{Svensson:2022kkj}, the $^3$H and $^3$He ground state
energies and radii at \NNNLO{} are also markedly worse compared to
\NNLO{}. This trend is a testament to the importance of inferring
\lec{}s in the \NN{}- and \NNN{}-sectors of \cheft{}
simultaneously~\cite{Carlsson:2015vda}.

We conclude, based on the inference of \NN{} \lec{}s made in
Ref.~\cite{Svensson:2022kkj}, that the discrepancies between
experimental low-energy \nd{} cross section data and theoretical
predictions are not due to the uncertainties stemming from the \lec{}
variability. Given the very narrow \pdf{s} for the \lec{s}, an
opposite finding would be a testament to a tremendous fine tuning of
scattering observables in the \NNN{} continuum relative to the \NN{}
continuum.

\subsection{The EFT truncation error}
\label{sec | eft error}
The truncation of the \cheft{} expansion used to describe the nuclear
interaction leads to a model discrepancy referred to as an EFT
truncation error. Following Ref.~\cite{Furnstahl:2015rha}, we assume
that the theoretical prediction at chiral order $\nu$ for some
observable $y$ can be written as
\begin{equation}
  y^{(\nu)}(\lecs;\vec{x}) = y_\text{ref}(\vec{x})\sum_{k=0}^{\nu} c_k(\lecs;\vec{x})Q^{\nu}(\vec{x})\:,
  \label{eq | eft expansion}
\end{equation}
where $\vec{x}$ denotes the kinematic variables $\Elab$ and $\thcm$
and $y_\text{ref}$ is a reference value for the observable in
question. This expression renders the expansion coefficients $c_k$
dimensionless quantities, which we also expect to be of natural size, i.e.,
$c_k \sim \mathcal{O}(1)$. We assume a \cheft{} expansion parameter of the form
\begin{equation}
  Q = \text{max}\left(\frac{q}{\Lambda_b},\frac{m_{\pi}}{\Lambda_b}\right),
  \label{eq | eft parameter}
\end{equation}
and set the \cheft{} breakdown scale to $\Lambda_b=600$ MeV as in
Ref.~\cite{Svensson:2022kkj} from where we also obtain the \lec{}
posteriors. We set the \cm{} momentum, $q$, according to the kinetic
energy, $\Elab$, of the incoming nucleon.
The \cheft{} truncation error, $\delta y_{\nu}$, is the expected
magnitude of the sum of contributions from terms beyond the order $\nu$.
Under the assumption of having
independent and normally distributed expansion coefficients, $c_k$, it
is shown in, e.g., Ref.~\cite{Melendez:2019izc}, that $\delta y_{\nu}$
is also normally distributed and given
by
\begin{equation}
  \delta y_{\nu} \sim \mathcal{N}\left( 0,y_\text{ref}^2\frac{Q^{2(\nu+1)}}{1-Q^2}\bar{c}^2\right),
  \label{eq | eft error}
\end{equation}
where $\bar{c}^2$ denotes the variance of the expansion coefficients.
Thus, knowing $\bar{c}^2$ enables us to quantify the (variance of the)
\cheft{} truncation error. For this purpose, we follow the procedure
of, e.g., Ref.~\cite{Svensson:2021lzs} and employ the root-mean-square
(RMS) value of order-by-order differences to estimate $\bar{c}^2$.
The order-by-order differences are computed from the mean
values of the \ppd{}s at each order $\nu$, thus averaging over a
possible \lec{} dependence.

We wish to compare the magnitude of the \cheft{} truncation error with
the theoretical error in $y$ stemming from the uncertainty about the
numerical values of the \lec{}s. Let us take the differential cross
section at $\Elab=12$ MeV as an example and inspect it
closer. Limiting ourselves to this low value of $\Elab$, the effect of
\NNNF{}s are expected to be small~\cite{Witala:1998ey}. Therefore, we
retain the expansion in Eq.~\eqref{eq | eft expansion} and use
Eq.~\eqref{eq | eft error} to quantify the \cheft{} truncation
error. We set $y_\text{ref}$ to the \LO{} prediction. At this
scattering energy, we also have $Q=m_{\pi}/\Lambda_b\approx 0.23$. An
RMS estimate from the expansion coefficients at $\thcm=30,\:90$ and
$150$ degrees (omitting \LO{} results due to their role in the
definition of $y_\text{ref}$) gives $\bar{c}=14.8$. This is a fairly
unnatural value which arises from an oscillating convergence when
including higher chiral orders.

The \ppd{}s due to the \lec{} variabilities and the \cheft{}
truncation errors are compared in Fig.~\ref{fig | ppd prob
  comparison}.
\begin{figure}[t!]
  \centering
  \includegraphics[width=\columnwidth]{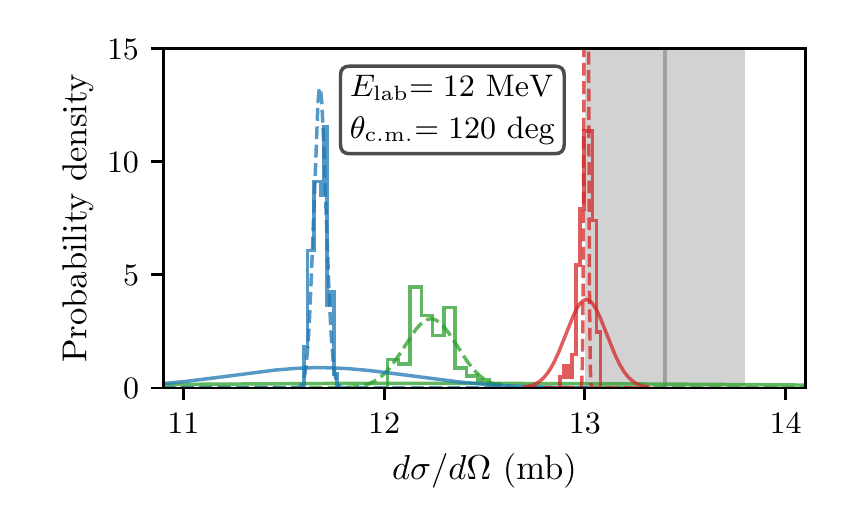}
  \caption{Comparison of the \ppd{} due to \lec{} variability and an
    estimate of the normally distributed \cheft{} truncation error for
    the differential cross section for \NLO{} (green), \NNLO{} (blue),
    and \NNNLO{} (red). The \cheft{} truncation is shown for two
    different variances; $\bar{c}=14.8$ (solid line) and $\bar{c}=1$
    (dashed line). To set the absolute scale, we included the
    experimental measurement (gray) from Ref.~\cite{Schwarz:1983rlz}}
  \label{fig | ppd prob comparison}
\end{figure}
Clearly, the truncation error (solid lines) is typically much greater
than the error due to the uncertain values of the \NN{} \lec{}s
(histograms) up to an including \NNLO. Therefore, we find it
unnecessary to account for a possible \lec{} variability in the expansion
coefficients in Eq.~\eqref{eq | eft expansion}. At \NNNLO{}, the two
errors are becoming comparable. However, at this order, both of the
errors are tiny, $\lesssim 4\%$, on an absolute scale. In fact, they
are both smaller than typical experimental errors, indicated as
the gray area in Fig.~\ref{fig | ppd prob comparison}. In addition to
the RMS estimate of $\bar{c}^2$ we also show the truncation error
(dashed line) based on a naturalness assumption where we set
$\bar{c}^2=1$. In this limit, the two errors become comparable for
this observable already at \NLO.

At higher energies, we see in Fig,~\ref{fig | ppd dsg} that the predictions at \NNNLO{} deviates from the ones at \NLO{} and \NNLO{}. When analyzing the truncation errors at $\Elab = 36$ MeV, we obtain $\bar{c}= 65.1$, which signals the presence of an unnaturally large contribution in the \cheft{} expansion. We find that omitting the shift between \NNLO{} and \NNNLO{} has a significant impact and yields a more reasonable value of $\bar{c}=15.1$. Doing the same at $\Elab = 12$ MeV yields $\bar{c}=11.3$, i.e., a relatively small change from when including the shift. The truncation error is expected to increase with the on-shell energy, and thus it should become greater than the \lec{} uncertainty for $\Elab > 12$ MeV, but we leave a more detailed study for future work.

\subsection{Spin-polarization observables}
\label{sec | polarization}
There are many different possibilities to form observables related to
spin-polarization in the initial and/or final states of the \Nd{}
reactants~\cite{Ohlsen:1972zz}. The fine details of the angular dependence of these
observables can depend sensitively on the spin structure of the \NN{}
and \NNN{} interactions. A well-known example is the low-energy vector
analyzing power $A_y$. This observable depends sensitively on the
$^3$P partial waves of the
\NN-interaction~\cite{Huber:1998hu,Margaryan:2015rzg}. There are
indications that it also depends sensitively on parts of the
subleading \NNNF~\cite{Epelbaum:2019zqc}. It has turned out to be very
challenging to reproduce the experimental data for this observable at
laboratory scattering energies $\Elab \lesssim$ 30
MeV~\cite{Gloeckle:1995jg,Weisel:2015min}.

\begin{figure}[ht!]
  \centering
  \includegraphics[width=\columnwidth]{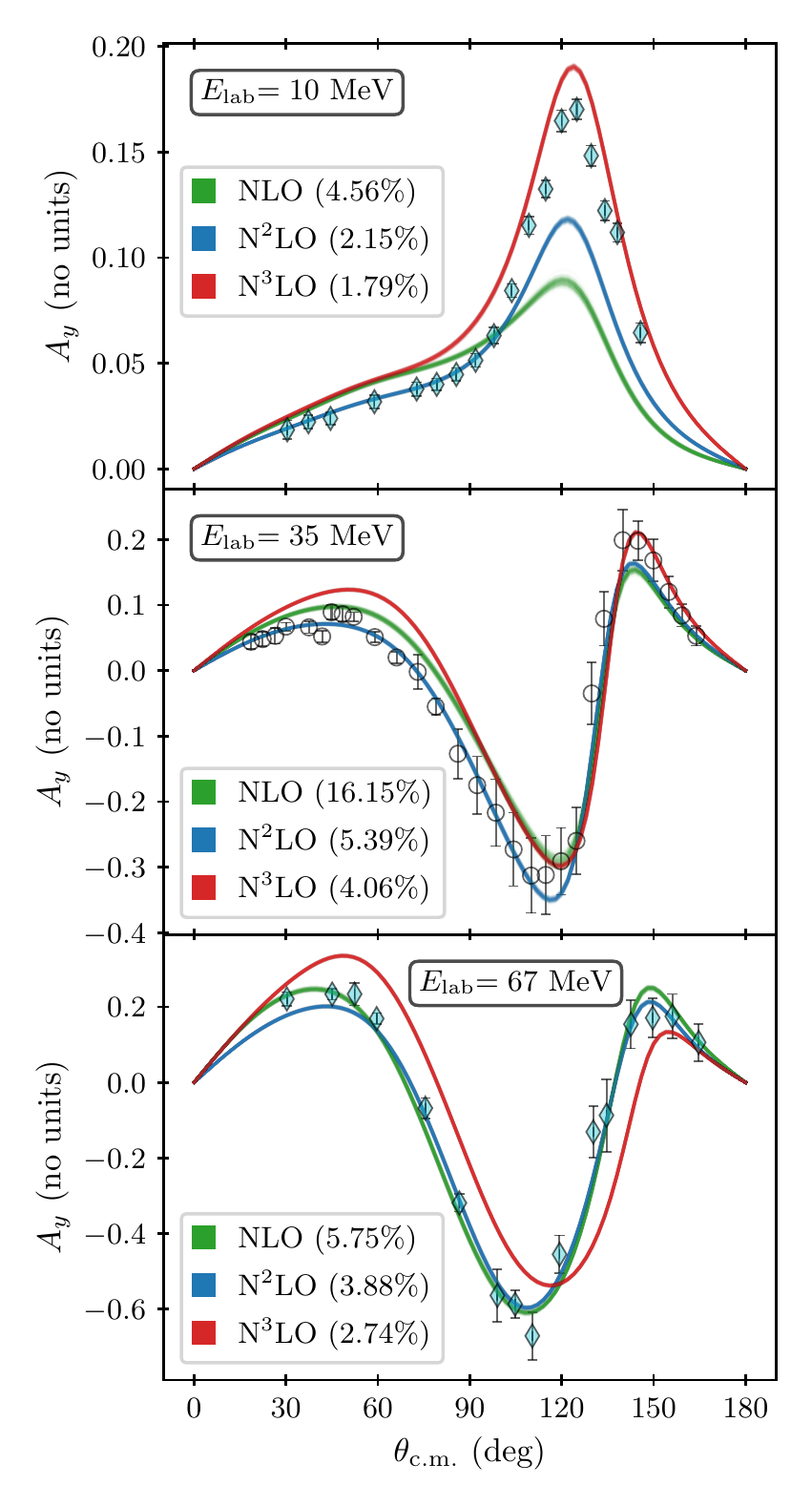}
  \caption{The \ppd{} of the \nd{} neutron analyzing power $A_y(n)$
    at $\Elab=10,\:35$, and 67 MeV up to \NNNLO{} in the \NN{}
    interaction. The legends display the average 90\% credible
    intervals (\hpdi{}), see
    text for details. The experimental data (markers) are retrieved
    from the EXFOR database (at $\Elab=10$ and 66.6 MeV) and
    Ref.~\cite{Bunker:1968omc} (at $\Elab=35$ MeV). Diamond (cyan) and
    circle (colorless) markers represent \nd{} and \pd{} cross sections, respectively.}
  \label{fig | ppd Ay}
\end{figure}

Given the possibly fine-tuned nature of $A_y$, it is particularly
interesting to quantify the \ppd{} due to the variability in the \NN{}
\lec{}s of \cheft{}. In Fig.~\ref{fig | ppd Ay}, we show the \ppd{}s
for $A_y$ at \NLO, \NNLO, and \NNNLO{} as well as the average 90\%
credibility intervals. At $\Elab=10$ MeV we do not reproduce the
experimental data at any chiral order. We note that the \NNNLO{}
calculation appears to improve the description of the data at the
polarization maximum. However, the low-angle description is markedly
worse compared to the result at \NNLO{}. For $\Elab \approx 35-67$ MeV 
it appears sufficient to use \NN-only forces at
\NNLO{} to describe presently available data.

It is clear that the variability due to the \lec{}s inferred from
\NN{} data does not give rise to any significant uncertainty nor does
it explain discrepancies between theory and data. We refrain from
quantifying the \cheft{} truncation error for this observable since our calculation omits
\NNNF{}s, which may very well play a significant role in explaining
the low-energy $A_y$ values. Nevertheless, a crude estimate to account
for the \cheft{} truncation error with missing \NNNF{}s can be obtained by
pulling out factors of $Q$ in Eq.~\eqref{eq | eft error}, starting at
\NNLO{}~\cite{LENPIC:2015qsz}. We found that this procedure induced rather large
\cheft{} uncertainties that covered the experimental data at all orders.

As for the remaining spin-polarization observables, their \NN{} \ppd{}s
exhibit similar patterns and widths as presented above for the
differential cross section and $A_y(n)$, i.e., the vastly dominating source
of uncertainty is the \cheft{} truncation error, at least below \NNNLO{}.

\section{Summary and outlook}
\label{sec | summary and outlook}
We sampled the \ppd{s} for the \nd{} differential cross section
$d\sigma/d\Omega$ at $\Elab = 12$, 36, and 65 MeV scattering
energy, and neutron analyzing power $A_y(n)$ at $\Elab = 10$, 35, and
67 MeV. The underlying samples from the \lec{} posterior were obtained
from a previous analysis of \NN{}
data~\cite{Svensson:2022kkj}. The \hmc{} algorithm used in
that analysis
yields virtually uncorrelated samples which we find most
likely persists for the elastic \nd{} observables.  The main
conclusion from this work is that the uncertainty about \NN{} \lec{}s,
when conditioned on \NN{} scattering data and uncorrelated estimates
of the \cheft{} truncation errors, does not entail significant
uncertainties in the low-energy \nd{} continuum. Although we only show
results for selected observables, we find them to be representative of
all elastic \nd{} scattering observables, at least for $\Elab \lesssim
67$ MeV.

When compared with estimates of the \cheft{} truncation error, we find
that the uncertainty stemming from the numerical values of the \NN{} \lec{}s
are negligible, at least up to (and including) \NNLO{} in Weinberg
power counting. At \NNNLO{}, the width of the \ppd{} and the
credible interval of the truncation error are starting to become
comparable. However, these uncertainties are very small and, in fact,
are comparable to typical experimental errors.

In this work we have not quantified the errors due to having a finite
number of wave-packets in the \wpcd{} method. Instead, we extrapolated
all results to $\nwp=150$ and relied on the fact that the widths and
shapes of all studied \ppd{}s remain the same when using fewer
wave-packets, i.e., $\nwp=50$ and 75. Future work should be dedicated to
understanding the scaling of the \wpcd{} method-error with respect to
the discretization of the continuum.

Throughout our analysis, the \ppd{}s were conditioned on \NN{}
scattering data. For the predicted differential cross section, we find reasonable
agreement with experimental \Nd{} scattering data. The same observation was made for many polarization observables, not shown explicitly in this paper. However, less accuracy is observed in the low-energy $A_y$ analyzing power. A natural next step
would therefore be to simultaneously infer the \NN{} and \NNN{} \lec{}s from
\NN{} plus \Nd{} scattering data. This would shed more light on the
necessity of including \NNNF{}s to explain this data.

The inference of \lec{}s in \cheft{} is not restricted to use only scattering observables. In fact, any low-energy nuclear data can be utilized (and will be relevant given that it has a high information content). On the other hand,
the abundant sets of experimentally measured
\NN{}~\cite{NavarroPerez:2013mvd,NavarroPerez:2013usk},
\piN{}~\cite{Workman:2012hx}, and
\Nd{}~\cite{Kalantar-Nayestanaki:2011rzs} scattering cross sections
provide data where theoretical predictions do not rely on many-body
interactions beyond \NNNF{}s. In addition, a scattering cross section
can be tied to a well-defined (external) momentum, providing a clear
interpretation of the soft scale entering the \cheft{} expansion
parameter $Q$ and the associated truncation error. This identification of a
soft scale is more ambiguous in bound states of nuclear many-body systems.

A Bayesian analysis of \lec{}s in \cheft{} conditioned on \Nd{} data
requires efficient solutions to the AGS equations. Indeed, traversing
larger domains of the multi-dimensional \lec{} parameter-spaces would
require orders of magnitude more samples than what we employed in
this work. Fortunately, recent advances in model reduction
methods~\cite{Melendez:2022kid}, utilizing singular value
decomposition~\cite{Tichai:2022mqn} and eigenvector
continuation~\cite{Frame:2017fah,Konig:2019adq,Zhang:2021jmi} methods,
show great promise in delivering accurate and fast solutions to the
Faddeev equations. Some of these methods appear compatible with our
existing implementation for solving the AGS equations with the \wpcd{}
method.

\section*{Acknowledgments}
This work was supported by the European Research Council (ERC) under
the European Unions Horizon 2020 research and innovation programme
(Grant agreement No.\ 758027). The work of C.\ F.\ was supported by the Swedish
Research Council (dnr.~2017-04234 and 2021-04507). The computations and data handling were
enabled by resources provided by the Swedish National Infrastructure
for Computing (SNIC), partially funded by the Swedish Research Council
through grant agreement no.\ 2018-05973.

\bibliography{bibliography}
\end{document}

%% file: preamble.tex
\usepackage{amsmath}
\usepackage{amssymb}
\usepackage{amsfonts}
\usepackage{mathtools}
\usepackage{bbm}
\usepackage{import}
\usepackage{graphicx}
\usepackage{fancyhdr}
\usepackage{tikz}
\usepackage{array}
\usepackage{units}
\usepackage{courier}
\usepackage{slashed}
\usepackage{bm}
\usepackage{dsfont}
\usepackage{epigraph}
\usepackage{verbatim}
\usepackage{listings}
\usepackage{comment}
\usepackage{multirow}
\usepackage{wrapfig} 
\usepackage{hyperref}
\usepackage{enumerate}

\usepackage[textsize=normalsize]{todonotes}

\newcounter{topic} 

\newcommand{\NNNLO}{N$^3$LO}
\newcommand{\NNLO}{N$^2$LO}
\newcommand{\NLO}{NLO}
\newcommand{\LO}{LO}

\newcommand{\cm}{c.m.}

\newcommand{\wpcd}{WPCD}

\newcommand{\pdf}{PDF}
\newcommand{\hmc}{HMC}
\newcommand{\hpdi}{HPDI}
\newcommand{\map}{MAP}

\newcommand{\cheft}{$\chi$EFT}
\newcommand{\lec}{LEC}
\newcommand{\lecs}{\ensuremath{\vec{\alpha}}}
\newcommand{\NN}{$NN$}
\newcommand{\NNN}{$NNN$}
\newcommand{\Nd}{$Nd$}
\newcommand{\piN}{$\pi N$}

\newcommand{\NNNF}{3NF}
\newcommand{\ppd}{PPD}
\newcommand{\pr}{\ensuremath{\text{pr}}}
\newcommand{\nn}{$nn$}

\newcommand{\nd}{$nd$}
\newcommand{\pd}{$pd$}

\newcommand{\Elab}{\ensuremath{E_\text{Lab}}}
\newcommand{\thcm}{\ensuremath{\theta_\text{c.m.}}}

\newcommand{\nwp}{N_{\text{WP}}}